\documentclass[preprint,12pt,sort&compress]{elsarticle}

\usepackage{lipsum}
\usepackage{xcolor,graphicx}
\usepackage{dcolumn}
\usepackage{bm}
\usepackage{subcaption}
\usepackage{booktabs}
\usepackage{amsmath,amssymb}
\usepackage[colorlinks,linkcolor=blue]{hyperref}
\usepackage{mathtools, cuted}
\usepackage[T1]{fontenc} 

\journal{Physics of the Dark Universe}

\begin{document}

\begin{frontmatter}



\title{Dynamical dark energy from spacetime-symmetry breaking - late-time behaviour and phantom crossing}


\author[1,2]{Nils A. Nilsson}\ead{nilsson@ibs.re.kr}
\affiliation[1]{Organization={Cosmology, Gravity and Astroparticle Physics Group, Center for Theoretical Physics of the Universe,
Institute for Basic Science}, addressline={Daejeon 34126, Korea}}
\affiliation[2]{organization={SYRTE, Observatoire de Paris, Universit\'e PSL, CNRS, Sorbonne Universit\'e,
LNE},
            addressline={61 avenue de l'Observatoire}, 
            city={Paris},
            postcode={75014}, 
            country={France}}

\begin{abstract}
We investigate the late-time cosmological dynamics in a simple case of explicit spacetime-symmetry breaking. By expanding in a small symmetry-breaking coefficient we are able to write the Friedmann equations as $\Lambda$CDM + dynamical dark energy, which we show contains logarithmic dependence of the scale factor. We find that the dark energy equation of state displays divergencies and phantom behaviour for certain values of the symmetry-breaking coefficient, where the NEC is also broken.
We discuss the adiabatic sound speed of dark energy and compare the model to current constraints using the Chevallier-Polarski-Linder parametrisation. Remarkably, although the constraints on the same symmetry-breaking coefficient from e.g. gravitational-wave propagation are orders of magnitude stronger than what we obtain in this paper, we are able to cut those constraints, which are more or less symmetric around zero, in half by showing that same coefficient must be negative (or zero) if one wishes to keep the NEC intact.
\end{abstract}




\end{frontmatter}




\section{\label{sec:intro}Introduction}
The accelerating expansion of the Universe was first discovered using type-Ia supernovae \cite{SupernovaSearchTeam:1998fmf,SupernovaCosmologyProject:1998vns}, and was awarded the Nobel prize in physics in 2011. Since then, significant effort has been put towards revealing the microphysics responsible for the acceleration, which to this day is not fully understood; this has lead to the term Dark Energy (DE). In the standard $\Lambda$ Cold-Dark-Matter ($\Lambda$CDM) model, the effects of DE are described through the cosmological constant $\Lambda$, which has negative pressure and becomes dominant once other cosmological fluids have decayed sufficiently, causing the acceleration. There is significant disagreement in the value of the cosmological constant: the difference between values obtained from the Cosmic Microwave Background \cite{Planck:2018vyg} and quantum field theory calculations of the vacuum energy currently lies around $55$ orders of magnitude\footnote{This differs from the most quoted discrepancy of 122 orders of magnitude due to details of renormalisation as described in \cite{Martin:2012bt}.}, which is known as the cosmological constant problem \cite{Martin:2012bt}; within the $\Lambda$CDM model, DE makes up around $68\%$ of the energy content of the Universe. As with other cosmic fluids, DE can be described using the barotropic index or equation of state parameter $w$ through $p = w\rho$, where $p$ and $\rho$ is the pressure and energy density, respectively. In the case of a cosmological constant, the equation of state parameter is exactly $w=-1$, but for more general models, $w$ may be a function of redshift. In addition to the cosmological constant problem, there is also an issue of fine-tuning of initial conditions known as the coincidence problem \cite{Zlatev:1998tr}. In light of these problems and tensions, it seems clear that new physics is needed. One may also wish to consider whether the issue lies with the $\Lambda$CDM model itself, and whether there exist some alternative in which these tensions are resolved \cite{Keeley:2022ojz,Krishnan:2021dyb}; if this is the case, there would be no need for new microphysics.

In order to address these outstanding issues, a number of Effective-Field Theories (EFT's) have been proposed throughout the years, usually attempting to replace the cosmological constant with a dynamical scalar field responsible for the effects of DE; amongst these EFT's, the most widely known are the quintessence \cite{Ratra:1987rm,Wetterich:1987fm} and k-essence models \cite{Armendariz-Picon:1999hyi,Chiba:1999ka,SupernovaSearchTeam:1998fmf}, but many others exist\footnote{See for example \cite{Frusciante:2019xia} for a review of DE EFT's.}. DE with $w<-1$ is known as {\it phantom dark energy}, the energy density of which increases with time (i.e. it has strongly negative pressure, and thus propagates against the direction of momentum) \cite{Ludwick:2017tox,Dabrowski:2003jm}. If this is actually the case, our Universe may eventually end up in one of several possible future singularities \cite{Nojiri:2005sx}. We may obtain a phantom fluid by reversing the sign of the kinetic term of a scalar field Lagrangian, but it also shows up naturally in certain higher-order theories of gravity \cite{Pollock:1988xe}, Brans-Dicke theories, and scalar-field theories with non-minimal coupling \cite{Torres:2002pe}. Generally, phantom fields exhibit a number of undesirable features, such as classical or quantum instabilities \cite{Buniy:2006xf}, anisotropy and superluminal propagation \cite{Dubovsky:2005xd}, or the lack of a Lorentz-invariant vacuum \cite{Arkani-Hamed:2003pdi,Nicolis:2008in}. In theories where Lorentz invariance is allowed to be broken, it may however be possible to render superluminal modes and instabilities unobservable \cite{Rubakov:2006pn}. On the other hand, it has been shown that DE EFT's with $w>-1$ in the local Universe generally lead to determinations of the Hubble constant which are lower than that of $\Lambda$CDM \cite{Banerjee:2020xcn,Lee:2022cyh}, thus exacerbating the mismatch of the Hubble parameter as measured with local probes as compared to its cosmological value, known as the Hubble tension (see for example \cite{DiValentino:2021izs,Riess:2020fzl,Freedman:2021ahq,Planck:2018vyg}).

It has been proposed in the literature that theories which break the foundational symmetries of General Relativity (GR) may provide solutions to some of the current cosmological puzzles, including the cosmological constant problem. For example, it was proposed in \cite{Blas:2011en} that dark energy may emerge naturally as a Goldstone field of a broken symmetry in the context of khronometric theories, a notion which was later tested in \cite{Audren:2013dwa}. A related approach is that of Ho\v{r}ava-Lifshitz gravity, which breaks Lorentz symmetry explicitly and which has been shown to contain dynamical DE with a phantom regime (for certain parameter values) \cite{Park:2009zr, Saridakis:2009bv,DiValentino:2022eot}. Therefore, we investigate in this paper the DE properties of a simple case of {\it explicit spacetime-symmetry breaking} in the form of a correction to the Einstein-Hilbert action \cite{ONeal-Ault:2020ebv}. This cosmological solution was found using a generic EFT framework used for testing spacetime symmetries in all sectors of the Standard Model as well as gravity \cite{Colladay:1996iz,Colladay:1998fq, Kostelecky:2003fs}, which has been extensively studied in the past decades (see \cite{Kostelecky:2008ts} for an annually updated list of constraints). On the level of cosmology, this EFT has been used to study inflation \cite{Bonder:2017dpb, Nilsson:2022mzq}, background evolution \cite{Reyes:2022dil,ONeal-Ault:2020ebv}, the Hubble parameter tension \cite{Khodadi:2023ezj}, metric anisotropies \cite{Maluf:2021lwh}, and more. In weak gravity, constraints on the EFT coefficients have been found using solar-system tests \cite{Bailey:2006fd,Hees:2013iv,LePoncin-Lafitte:2016ocy}, short-range gravity \cite{Long:2014swa,Shao:2016cjk,Shao:2018lsx,Bailey:2022wuv}, pulsar tests \cite{Shao:2018vul,Shao:2014oha}, gravitational waves \cite{ONeal-Ault:2021uwu,Wang:2021ctl,Liu:2020slm}, and many more.

This paper is organised as follows: in Section~\ref{sec:theory} we introduce the field theory and the resulting cosmology; in Section~\ref{app:backevol} we isolate the effects of the resulting dynamical DE and study its properties; we discuss our results and conclude in Section~\ref{sec:disc}. Throughout this paper we use a standard flat FLRW cosmology with mostly-plus signature.

\section{Cosmology with explicit spacetime-symmetry breaking\label{sec:theory}}
We can write the Lagrange density using the vierbein formalism as
\begin{equation}
\begin{aligned}
    \mathcal{L} = \frac{e}{2\kappa}[R-2\Lambda&+a^{\lambda\mu\nu}T_{\lambda\mu\nu}+b^{\kappa\lambda\mu\nu}R_{\kappa\lambda\mu\nu}].
\end{aligned}
\end{equation}
where $R$ is the curvature scalar, $\Lambda$ is the cosmological constant, $e$ is the determinant of the vierbein, $R_{\kappa\lambda\mu\nu}$ is the Riemann tensor, and $T_{\lambda\mu\nu}$ is the torsion tensor. Also present are the quantities $a^{\lambda\mu\nu}$ and $b^{\kappa\lambda\mu\nu}$, which are the nondynamical background fields which transform as scalars under so-called particle rotations \cite{Kostelecky:2003fs}, and the above action therefore explicitly breaks particle diffeomorphisms whilst still respecting observer general coordinate invariance. As a consequence of the broken diffeomorphisms at the level of the action, the adherence of the theory to the traced Bianchi identity is no longer guaranteed, and has to be imposed by hand. We note that when working with some background tensor $k_{\mu\nu}$ in the spacetime frame, using the vierbein to transform $k_{\mu\nu}$ to the locally Lorentz frame as $k_{ab}=e^{\mu}_{~a}e^{\nu}_{~b}k_{\mu\nu}$ results in a {\it different theory} compared to using $k_{\mu\nu}$ to contract directly with fields in the local frame.

In the Riemannian limit, the torsion vanishes and we can express the theory using the metric tensor as 
\begin{equation}\label{eq:lagr}
    \mathcal{L}=\frac{\sqrt{-g}}{2\kappa}[R-2\Lambda+b^{\kappa\lambda\mu\nu}R_{\kappa\lambda\mu\nu}],
\end{equation}
where the symmetry-breaking term can be decomposed according to the symmetry properties of the Riemann tensor as
\begin{equation}
    b^{\kappa\lambda\mu\nu}R_{\kappa\lambda\mu\nu} = \underbrace{-uR + s^{(T)}_{\mu\nu}R^{(T)\mu\nu}}_{=s^{}_{\mu\nu}R^{\mu\nu}}+t^{\kappa\lambda\mu\nu}C_{\kappa\lambda\mu\nu},
\end{equation}
where $R^{(T)\mu\nu}$ denotes the trace-free Ricci tensor and $C_{\kappa\lambda\mu\nu}$ the Weyl tensor. The term $-uR$ represents the trace part of the second term. In order to keep the equations of motion more tractable, we will consider the case where $t^{\kappa\lambda\mu\nu}=0$ from now on.\footnote{Since we later restrict ourselves to the flat Friedmann-Lemaitre-Robertson-Walker metric, the Weyl tensor term will vanish due to conformal flatness, and the choice $t^{\kappa\lambda\mu\nu}=0$ does not affect the results.} Also, we reabsorb the scalar trace term into the two-tensor $s_{\mu\nu}$; this tensor explicitly breaks the spacetime symmetries and will be the main topic of study in this paper.

We arrive at the field equations by varying the action $S=\int d^4x \mathcal{L}$ using the Lagrange density $(\ref{eq:lagr})$, after which we find
\begin{equation}
    \begin{aligned}
        R_{\mu\nu}&-\tfrac{1}{2}g_{\mu\nu}R-\tfrac{1}{2}g_{\mu\nu}R^{\alpha\beta}s_{\alpha\beta}+2R_{(\mu}^{~\alpha}s_{\nu)\alpha}+\tfrac{1}{2}\Box s_{\mu\nu} \\&-\nabla_\alpha\nabla_{(\mu}s_{\nu)}^{~\alpha}+\tfrac{1}{2}g_{\mu\nu}\nabla_\alpha\nabla_{\beta}s^{\alpha\beta} = \kappa T_{\mu\nu},
    \end{aligned}
\end{equation}
where $\Box = \nabla_\lambda\nabla^\lambda$ is the covariant d'Alembertian, $T_{\mu\nu}$ is the stress-energy tensor for matter and dark energy, and parentheses denote symmetrisation of indices. Collecting all terms proportional to $s_{\mu\nu}$ to the right-hand side allows us to write the Einstein equations as
\begin{equation}\label{eq:gmunu}
\begin{aligned}
    G_{\mu\nu}&=\kappa T_{\mu\nu} + (T_s)_{\mu\nu} \\
    (T_s)_{\mu\nu} &= \tfrac{1}{2}g_{\mu\nu}R^{\alpha\beta}s_{\alpha\beta}-2R_{(\mu}^{~\alpha}s_{\nu)\alpha}-\tfrac{1}{2}\Box s_{\mu\nu} \\&+\nabla_\alpha\nabla_{(\mu}s_{\nu)}^{~\alpha}-\tfrac{1}{2}g_{\mu\nu}\nabla_\alpha\nabla_{\beta}s^{\alpha\beta},
\end{aligned}
\end{equation}
where $G_{\mu\nu}$ is the Einstein tensor and $(T_s)_{\mu\nu}$ represents the stress-energy tensor arising from the symmetry-breaking coefficients $s_{\mu\nu}$.

The conservation laws of the underlying action in the form of the traced Bianchi identities $\nabla_\mu G^{\mu\nu}=0$ must be satisfied for the theory to be self consistent. Applying $\nabla_\mu$ on both sides of Eq.~\eqref{eq:gmunu} gives

\begin{equation}
        \kappa\nabla_\mu T^\mu_{~\nu} = -\tfrac{1}{2}R^{\alpha\beta}\nabla_\nu s_{\alpha\beta}+R^{\alpha\beta}\nabla_\beta s_{\alpha\nu}+\tfrac{1}{2}s_{\alpha\nu}\nabla^\alpha R,
\end{equation}
which we can write as $\nabla_\mu[\kappa T^\mu_{~\nu} -(T_s)^\mu_{~\nu}]=0$. By demanding that the total right-hand side of the modified Einstein equations be conserved, i.e. {\it not} imposing the usual $\nabla_\mu T^{\mu}_{~\nu}=0$, we are modifying the cosmological evolution of the matter fields proportional to the coefficients of spacetime-symmetry breaking. It should be noted that if we had imposed the on-shell conservation of $T^{\mu}_{~\nu}$ and $(T_s)^{\mu}_{~\nu}$ separately, the resulting solution would have contained divergences and other pathological behaviour \cite{ONeal-Ault:2020ebv}.

From now on, we will restrict our attention to the case when only one component of the coefficient tensor is non-zero, so we choose the ansatz
\begin{equation}\label{eq:smunu}
s_{\mu\nu} =
    \begin{pmatrix}
        \mathcal{S} & 0 & 0 & 0 \\
        0 & 0 & 0 & 0 \\
        0 & 0 & 0 & 0 \\
        0 & 0 & 0 & 0 \\
    \end{pmatrix}.
\end{equation}
It can be shown \cite{ONeal-Ault:2021uwu} that the spatial parts of the Bianchi identities can be satisfied by assuming that the symmetry-breaking coefficient $\mathcal{S}$\footnote{We caution the reader that the coefficient $\mathcal{S}$ is referred to as $s_{00}$ in other works, e.g. \cite{ONeal-Ault:2021uwu,Nilsson:2022mzq}.} is spatially constant in the chosen coordinate system, and we will therefore adopt $\partial_i \mathcal{S}=0$ from now on; for simplicity, we will also impose that $\mathcal{S}$ is a constant w.r.t coordinate time $t$, i.e. $\partial_t \mathcal{S}=0$.

Introducing the flat Friedmann-Lemaitre-Robertson-Walker (FLRW) metric as
\begin{equation}
    ds^2 = -dt^2 +a(t)^2\left[dx^2+dy^2+dx^2\right],
\end{equation}
we obtain the following Friedmann equations
\begin{equation}\label{eq:fried2}
    \begin{aligned}
        \left(\frac{\dot{a}}{a}\right)^2 &= \frac{\kappa\rho}{3(1-\tfrac{3}{2}\mathcal{S})}+\frac{\kappa p \mathcal{S}}{(2-3\mathcal{S})(1-\mathcal{S})} \\
        \frac{\ddot{a}}{a} &= -\frac{\kappa(\rho+3p)}{6(1-\tfrac{3}{2}\mathcal{S})}
    \end{aligned}
\end{equation}
and when adopting a standard perfect-fluid stress-energy tensor $T^\mu_\nu={\rm diag}(-\rho,p,p,p)$, we find the $\nu=0$ Bianchi identities
\begin{equation}\label{eq:bianchi0}
\begin{aligned}
&\nabla_\mu(T_s)^\mu_{~0} = 6\mathcal{S}\frac{\ddot{a}}{a}\frac{\dot{a}}{a}+3\mathcal{S}\frac{\dddot{a}}{a}, \\
&\kappa\nabla_\mu T^\mu_{~0} = -\dot{\rho}-3\frac{\dot{a}}{a}(\rho+p).
\end{aligned}
\end{equation}
By plugging Eq.~\eqref{eq:fried2} into Eq.~\eqref{eq:bianchi0} and imposing $\nabla_\mu[\kappa T^\mu_{~\nu} -(T_s)^\mu_{~\nu}]=0$
we find a modified continuity equation of the form\footnote{See \ref{app:sanity} for a parallel derivation for the individual perfect-fluid constituents.}
\begin{equation}\label{eq:cont}
    \dot{\rho}+3Hf(\mathcal{S},w)\rho=0, \quad f(\mathcal{S},w)=\frac{2(1+w-\mathcal{S})}{2+\mathcal{S}(3w-2)}
\end{equation}
where where $H\equiv \dot{a}/a$ is the Hubble parameter and $f(\mathcal{S},w)$ is an auxiliary function. We can integrate the modified continuity equation to find
\begin{equation}\label{eq:aux}
    \rho \sim a^{-3f(w,\mathcal{S})},
\end{equation}
which we break into the constituent fluids as $\rho=\sum_i\rho_i$.
This modification leads to non-standard cosmological evolution\footnote{Although the equation of state parameters $w_x$ are still constant.} of radiation ($w=1/3$) and cosmological constant ($w=-1$), which can be seen by plugging in the corresponding values of the barotropic index into Eq.~(\ref{eq:cont})\footnote{A similar modification was found in the context of a type of interacting dark energy; see for example Eq.~12 in \cite{Bisnovatyi-Kogan:2023frj}.}; dust ($w=0$), and curvature ($w=-1/3$) receive no modification and evolve as usual (a plot of the auxiliary function can be see in Appendix C of \cite{Nilsson:2022mzq}).
In terms of the normalised energy densities $\Omega_r^0$ and $\Omega_\Lambda^0$, the evolution is modified as 
\begin{equation}\label{eq:omegamods}
    \Omega_r^0 a^{-4} \to \Omega_r^0 a^{-4x_r}, \quad \Omega_\Lambda^0 \to \Omega_\Lambda^0 a^{-x_\Lambda}, 
\end{equation}
where $x_r$ and $x_\Lambda$ are polynomials in $\mathcal{S}$ and read $x_r=(1-\tfrac{3}{4}\mathcal{S})/(1-\tfrac{1}{2}\mathcal{S})$, $x_\Lambda=-3\mathcal{S}/(1-\tfrac{5}{2}\mathcal{S})$. The change in the evolution is ''small``, since $|\mathcal{S}|$ must be much smaller than unity\footnote{As experiment has determined that Lorentz symmetry holds to very high precision.}; nevertheless, the symmetry breaking induces evolution in $\Lambda$ where previously there was none. An interesting phenomenological consequence of this modification might be its effect on the Hubble tension. Such a tension actually emerges naturally in symmetry-breaking models, as was first discussed in \cite{Carroll:2004ai}; however, as was shown in \cite{Khodadi:2023ezj}, the approach we take in this paper does not affect the present Hubble tension.
The Friedmann equations in Eq.~\eqref{eq:fried2} contains different scalings of known GR terms and also the presence of a pressure contribution which is not present in GR. Many of these scalings can be absorbed \cite{ONeal-Ault:2021uwu} into the definition of the density parameters of matter, radiation, and cosmological constant, and are thus unobservable. In the end, we can write the final Friedmann equations as
\begin{equation}\label{eq:fried}
    \begin{aligned}
        H^2& = H_0^2\left[\Omega_m^0 a^{-3}+\Omega_r^0 a^{-4x_r}+\Omega_\Lambda^0 a^{-x_\Lambda}\right],\\
        \dot{H}& +H=H_0\Big[\tfrac{1}{2}\Omega_m^0 a^{-3}-\Omega_r^0\frac{2(1-\mathcal{S})}{2-\mathcal{S}}a^{-4x_r}\\&+\Omega_\Lambda^0\frac{2(1-\mathcal{S})}{2-5\mathcal{S}}a^{-x_\Lambda}\Big],
    \end{aligned}
\end{equation}
where $H_0$ is the value of the Hubble parameter at the present time, and the quantities $\Omega_X^0$ denote the normalised densities for matter, radiation, and cosmological constant, respectively. 

\section{Dark energy}\label{app:backevol}
Spacetime symmetries have been tested with very high precision, and in the gravity sector, most constraints have been obtained in the spontaneous breaking case; all constraints using the present EFT can be found in \cite{Kostelecky:2008ts}. The weak gravity spontaneous-breaking analogue of the coefficient $\mathcal{S}$ (denoted $\bar{s}_{00}^{(0)}$ in \cite{Kostelecky:2008ts}) has received bounds from e.g. the speed of gravity measurement using GW170817 + GR170817A, which lead to $\bar{s}_{00}^{(4)}=(-20 \text{ to } +5)\cdot 10^{-15}$~\cite{LIGOScientific:2017zic,Kostelecky:2008ts}. $\mathcal{S}$ has also received bounds from GW propagation using the modified propagation of tensor modes, obtaining $-6\cdot 10^{-15}<\mathcal{S}<+7\cdot 10^{-16}$ \cite{Nilsson:2022mzq}. We note that this was obtained from the same explicit-breaking model under study in this paper, but where the metric was linearised (i.e. without splitting $\mathcal{S}$ into background + fluctuation as one would do in the spontaneous-breaking case).

Given the strong constraints on $\mathcal{S}$ as outlined above, we expand the Friedmann equations (\ref{eq:fried}) to second order in $\mathcal{S}$, after which they read
\begin{equation}\label{eq:friedexp}
    \begin{aligned}
        H^2&=H_0^2\Big[\Omega_m^0 a^{-3}+\Omega_r^0 a^{-4} + \Omega_\Lambda + \Omega_k^0 a^{-2}+\mathcal{S}(\Omega_r^0 a^{-4}\ln{a}\\&+3\Omega_\Lambda^0 \ln{a})+\mathcal{S}^2(\tfrac{1}{2}\Omega_r^0(\ln{a}+\ln{a}^2)a^{-4}+\tfrac{3}{2}\Omega_\Lambda^0(5\ln{a}+3\ln{a}^2))\Big] \\
        \dot{H}& +H=H_0^2\Big[-\tfrac{1}{2}\Omega_m^0 a^{-3}-\Omega_r^0 a^{-4}+\Omega_\Lambda^0 + \mathcal{S}(\tfrac{1}{2}\Omega_r^0(1-2\ln{a})a^{-4}\\&+\tfrac{3}{2}\Omega_\Lambda^0(1+2\ln{a}))+\mathcal{S}^2(\tfrac{1}{4}\Omega_r^0(1-2\ln{a}^2)a^{-4}+\tfrac{3}{4}\Omega_\Lambda^0(5\\&+16\ln{a}+6\ln{a}^2))\Big],
    \end{aligned}
\end{equation}
and we see that the effects of the modified continuity equation can be represented by standard $\Lambda$CDM cosmology plus a dynamical dark-energy term with logarithmic dependence of the scale factor; for example, we can write the first equation as 
\begin{equation}
H^2=H_0^2[\Omega_m^0 a^{-3}+\Omega_r^0 a^{-4} + \Omega_{\rm DE}(a)],
\end{equation}
where $\Omega_{\rm DE}(a)$ represents all symmetry-breaking terms along with the cosmological constant. 

We find the energy density $\rho_{\rm DE}$ and pressure $p_{\rm DE}$ of the dark energy as
\begin{equation}
    \begin{aligned}
        \rho_{\rm DE} &= \frac{H_0^2}{\kappa}\Big[\Omega_\Lambda^0+\mathcal{S}(\Omega_r^0 a^{-4}\ln{a}+4\Omega_\Lambda^0\ln{a})\\&+\mathcal{S}^2(\tfrac{1}{2}\Omega_r^0a^{-4}(\ln{a}+\ln{a}^2)+\tfrac{3}{2}\Omega_\Lambda^0(5\ln{a}+3\ln{a}^2))\Big]  \\
        p_{\rm DE} &= - \frac{H_0^2}{\kappa}\Big[3\Omega_\Lambda^0 + \mathcal{S}(\Omega_r^0a^{-4}+3\Omega_\Lambda^0-\Omega_r^0 a^{-4} \ln{a}\\&+9\Omega_\Lambda^{0}\ln{a})+\mathcal{S}^2\tfrac{1}{2}(\Omega_r^0a^{-4}+15\Omega_\Lambda^0+(\Omega_r^0 a^{-4}\\&+63\Omega_\Lambda^0)\ln{a}-(\Omega_r^0 a^{-4}-27\Omega_\Lambda^0)\ln{a}^2)\Big],
    \end{aligned}
\end{equation}
from which we obtain the dark-energy equation of state parameter $w_{\rm DE}=p_{\rm DE}/\rho_{\rm DE}$, which can be found in Appendix~\ref{app:longs}\footnote{Similar equations of state, with logarithmic dependence of the scale factor, were found in \cite{Kouwn:2012np} and \cite{Oikonomou:2019nmm}.}; it can easily be checked that $w_{\rm DE} \to -1$ as $\mathcal{S} \to 0$. We use a standard set of values for the cosmological parameters when generating the figures below:
    $\Omega_r = 10^{-4}, \quad \Omega_\Lambda = 0.7$.
    
\begin{figure}[h]
    \centering
    \includegraphics[scale=0.8]{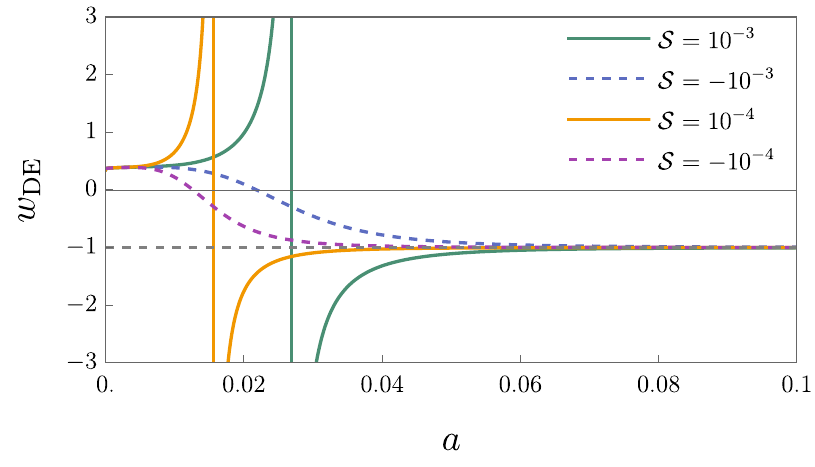}
    \caption{The dark-energy equation of state for different values of the coefficient $\mathcal{S}$. Negative values can be seen to ensure a smooth transition to $w_{\rm DE}\approx -1$.}
    \label{fig:wdea}
\end{figure}

For small values of $a$, the $w_{\rm DE}$ mimics that of radiation, with a formal limit $w_{\rm DE}\to 1/3$ as $a\to0$. As $a$ increases, we see that there exists a divergence when $\mathcal{S}$ is positive, after which $w_{\rm DE}$ settles down to a value close to minus one, i.e. almost pure cosmological constant. For negative values of $\mathcal{S}$, the transition in $w_{\rm DE}$ is smooth, and shows no divergent behaviour. We plot the behaviour of $w_{\rm DE}$ in Figure~\ref{fig:wdea} for different values of $\mathcal{S}$. 
In Figure~\ref{fig:wdea}, we see that reducing the value of $\mathcal{S}$ pushes the transition from a positive to negative equation of state to higher values of $a$, i.e. into the future. The same effect can be achieved by adjusting the value of $\Omega_\Lambda^0$: a increase (decrease) in this parameter pushes the transition to earlier (later) times. Adjusting the value of the radiation density has very similar effects, moving the transition point further into the future (by decreasing $\Omega_r^0$) or further into the past (by increasing $\Omega_r^0$). The only way to completely remove a divergent phantom crossing for $\mathcal{S}>0$ is to set $\Omega_r^0 \to 0$; interestingly, setting $\Omega_r^0$ to zero for $\mathcal{S}<0$ {\it introduces} a divergent crossing.

\begin{figure}[h]
    \centering
    \includegraphics[scale=0.8]{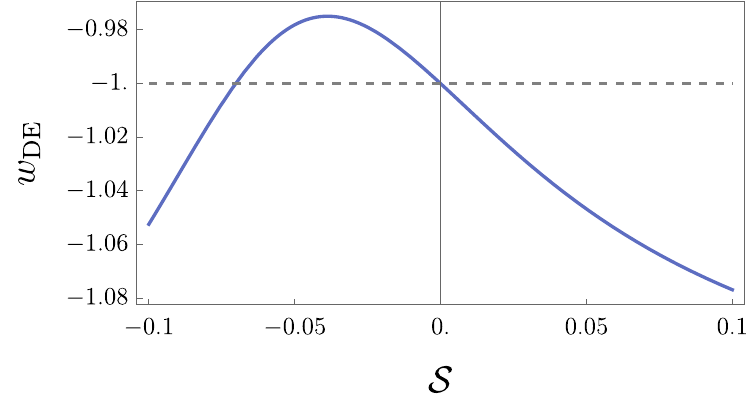}
    \caption{The dark energy equation of state as a function of the coefficient $\mathcal{S}$ with the scale factor fixed to $a=50$. A negative values of $\mathcal{S}$ is necessary in order to avoid phantom behaviour.}
    \label{fig:wdes00}
\end{figure}

Although not easily visible in Figure~\ref{fig:wdea}, $w_{\rm DE}$ only reaches minus one for certain values of the coefficient $\mathcal{S}$; in fact, for $\mathcal{S}<0$, which we have to choose if we want to avoid the divergences seen in Figure~\ref{fig:wdea}, $w_{\rm DE}$ is always greater than minus one, which can be seen in Figure~\ref{fig:wdes00}, where we have temporarily fixed $a=50$\footnote{Note here that since we have expanded in $\mathcal{S}$, only ``small'' values should be taken into account. In the plots, we exaggerate the magnitude of $\mathcal{S}$ in order to show the features more clearly.}.

\subsection{The CPL parametrisation and check of logarithmic corrections}\label{sec:CPL}
The Chevallier-Polarski-Linder parametrisation of the DE equation of state is one of the standard tools used to represent the unknown properties of DE in the late Universe \cite{Chevallier:2000qy,Linder:2002et}. For values of the scale factor close to the value at the present day ($a_0=1$), we may expand the equation of state for dark energy as
\begin{equation}
    w_{\rm DE} = w_0 + w_a (1-a) + w_b (1-a)^2+\hdots,
\end{equation}
and we find from the DE equation of state (Eq.~(\ref{eq:wDE})) that
    \begin{align}\label{eq:cplw0}
        w_0=&-1-\mathcal{S}\left(1+\frac{\Omega_r^0}{3\Omega_\Lambda^0}\right)-\mathcal{S}^2\frac{\Omega_r^0+15\Omega_\Lambda^0}{6\Omega_\Lambda^0} \\
        \label{eq:cplwa}w_a&= -\mathcal{S}\frac{8\Omega_r^0}{\Omega_\Lambda^0}-\mathcal{S}^2\frac{\Omega_r^0(\Omega_r^0+9\Omega_\Lambda^0)}{3(\Omega_\Lambda^0)^2},
    \end{align}
where in $w_a$ we have excluded terms of higher than second order in $\mathcal{S}$, which can be used to place direct limits on $\mathcal{S}$ from data. One of the more recent perturbation-level analyses \cite{Yang:2021flj} used a combination of the final Planck 2018 data release, Baryon Acoustic Oscillation measurements (BAO), and the Cosmic Distance Ladder (CDL) calibrated with Cepheid variable stars. The results revealed that to $1\sigma$, $w_0$ is distinctly negative ($w_0=-1$ and $w_a=0$ gives a pure cosmological constant), and that $w_a$ is consistent with zero. The exact values are given in Table~\ref{tab:1},
\begin{table}[h]
\centering
\resizebox{0.95\linewidth}{!}{%
\begin{tabular}{@{} l *4c @{}}
\toprule
 \multicolumn{1}{c}{Parameter from \cite{Yang:2021flj}} & Planck & Planck+BAO & Planck+CDL \\ 
\midrule
 $w_0$ & $-1.21^{+0.33}_{-0.60}<-0.37$ & $-0.67\pm0.32$ & $-0.89^{+0.32}_{-0.16}$ \\ 
 $w_a$ & $<-0.85<0.71$ & $-1.05^{+0.99}_{-0.77}$ & $<-1.04<0.47$\\ 
 $\Omega_m^0$ & $0.215^{+0.024}_{-0.078}$ & $0.335\pm0.029$ & $0.261^{+0.010}_{-0.011}$\\
 $\Omega_r^0\cdot 10^5$ & $6.34^{+0.71}_{-0.23}$ & $9.88\pm0.85$ & $7.69^{+0.29}_{-0.32}$ \\ 
 $H_0$ [km s$^{-1}$ Mpc$^{-1}$] & $84^{+15}_{-7}>63$ & $65.5^{+2.4}_{-3.2}$ & $74.1\pm1.4$\\
 \midrule
 $\mathcal{S}$ using Eq.~\eqref{eq:cplw0} & $(-0.63, +0.81)$ & $(-0.65,0.01)$ & $(-0.43, 0.05)$ \\
 \bottomrule
 \end{tabular}}
 \caption{$1\sigma$ constraints on the pertinent cosmological parameters from \cite{Yang:2021flj} with the resulting bounds on $\mathcal{S}$ from Eq.~\eqref{eq:cplw0} at linear order in $\mathcal{S}$. We can estimate $\Omega_\Lambda^0=1-\Omega_m^0-\Omega_r^0$.}
 \label{tab:1}
\end{table}
We also consider results obtained using a background analysis presented in \cite{Escamilla-Rivera:2021boq}, using the Pantheon catalogue of Supernovae Type Ia (Pantheon), Cosmic Chronometers (CC), as well as Gravitational-Wave events (GW) in the GWTC-1 and GWTC-2 catalogues from the LIGO/Virgo collaborations; these parameters along with our derived constraints on $\mathcal{S}$ can be seen in Table~\ref{tab:2}. When deriving these constraints, we have also used the value of $\Omega_r^0$ as $\Omega_r^0 = \Omega_m^0/(1+z_{\rm eq})$, where $z_{\rm eq}$ is the redshift at the matter-radiation equality, for which we adopt the Planck 2018 central value of $z_{\rm eq}=3402$ \cite{Planck:2018vyg}. 
\begin{table}[h]
\centering
\resizebox{0.95\linewidth}{!}{%
\begin{tabular}{@{} l *4c @{}}
\toprule
 \multicolumn{1}{c}{Parameter from \cite{Escamilla-Rivera:2021boq}} & GW & Pantheon+GW & Pantheon+GW+CC \\ 
\midrule
 $w_0$ & $-1.0^{+2.7}_{-3.0}$ & $-1.19\pm0.17$ & $-0.948^{+0.052}_{-0.074}$ \\ 
 $w_a$ & $0.2\pm2.9$ & $0.01^{+1.8}_{-0.67}$ & $0.60^{+0.47}_{-0.18}$\\ 
 $\Omega_m^0$ & $0.27^{+0.22}_{-0.12}$ & $0.309^{+0.12}_{-0.044}$ & $0.169^{+0.079}_{-0.031}$\\
 $H_0$ [km s$^{-1}$ Mpc$^{-1}$] & $74.00^{+0.12}_{-0.18}$ & $73.36\pm0.0042$ & $73.34^{+0.0032}_{-0.0028}$\\
 \midrule
 $\mathcal{S}$ using Eq.~\eqref{eq:cplw0} & $(-2.70,3.00)$ & $(0.02, 0.36)$ & $(-0.10, 0.02)$ \\
 \bottomrule
 \end{tabular}}
 \caption{$1\sigma$ constraints on the pertinent cosmological parameters from \cite{Escamilla-Rivera:2021boq} with the resulting bounds on $\mathcal{S}$ from Eq.~\eqref{eq:cplw0} at linear order in $\mathcal{S}$. We can estimate $\Omega_\Lambda^0=1-\Omega_m^0-\Omega_r^0$ and $\Omega_r^0 = \Omega_m^0/(1+z_{\rm eq})$, where $z_{\rm eq}$, where $z_{\rm eq}=3402$ is the Planck 2018 central value for the matter-equality redshift \cite{Planck:2018vyg}.}
 \label{tab:2}
\end{table}

Since Eq.~\eqref{eq:cplw0} and \eqref{eq:cplwa} are quadratic in $\mathcal{S}$, some of the parameter values in Table~\ref{tab:1} and \ref{tab:2} result in complex solutions for $\mathcal{S}$. To mitigate this, we consider only the terms in these equations which are {\it linear} in $\mathcal{S}$, remembering that this coefficient should be small. When doing this we find that Eq.~\eqref{eq:cplwa} leads to constraints on the order of $\pm10^3$, far outside the radius of convergence of the linear expansion we have made. Therefore, we present in Table~\ref{tab:1} and \ref{tab:2} the results from Eq.~\eqref{eq:cplw0} to linear order in $\mathcal{S}$. We obtain the tightest constraint from the Pantheon+GW+CC combination of data from the background analysis in \cite{Escamilla-Rivera:2021boq}, and the least competitive from GW only, at $1\sigma$. Interestingly, since the Pantheon+GW data combination prefers $w_0\neq 1$ at $1\sigma$, we obtain a constraint on $\mathcal{S}$ which excludes zero. This is in sharp contrast to the bound obtained in \cite{Nilsson:2022mzq} from the propagation speed of gravitational waves, which yielded $-6\cdot 10^{-15}<\mathcal{S}<+7\cdot 10^{-16}$.

\subsubsection{Check of logarithmic corrections}
It was reported in \cite{Efstathiou:1999tm} that a logarithmic relationship between the scale factor and the barotropic index provides a good fit to Supernovae Type Ia and Cosmic Microwave Background data. In this paper, the authors used tracker solutions to explore the parameter space and arrived at $w = w_Q + \alpha \ln{a}$, with limits on the parameters being approximately $w_Q \sim (-1,-0.4)$, $\alpha \sim (-0.05,-0.14)$. Our expression for $w$ cannot be put into the form in \cite{Efstathiou:1999tm} with constant parameters; instead, we can obtain tracker-like solutions where the parameters $w_Q$ and $\alpha$ vary slowly with the scale factor. By expanding our expression for $w$ in \ref{app:longs} in $\ln{a}$, we obtain the following expressions for $w_Q$ and $\alpha$ to first order in $\mathcal{S}$
\begin{equation}
    \begin{aligned}
        w_Q =& \frac{4\Omega_r^0 \mathcal{S}}{6\Omega_r^0 \mathcal{S}+3\Omega_\Lambda^0a^4(1+6\mathcal{S})} \\
        \alpha =& -\tfrac{10\Omega_r^0\mathcal{S}-3\Omega_\Lambda^0 a^4(1+10\mathcal{S})}{9\Omega_r^0+3\Omega\Lambda^0a^4(1+9\mathcal{S})},
    \end{aligned}
\end{equation}
both of which vary slowly in $a$. By plugging in the same values for the cosmological parameters as in Section~\ref{sec:CPL}, we find that $w_Q$ stays within the quoted limit of $(-1,-0.4)$ for $\mathcal{S}<0$, even at $a \to 1$, the original definition in \cite{Efstathiou:1999tm}; however, we also see that $\alpha$ is three orders of magnitude too large ($\sim 10^{-5}$ versus the quoted $10^{-2}$), but we notice that it receives that correct sign for $\mathcal{S}<0$. Although it seems that a logarithmic correction cannot accurately capture the features of the present model, we do reaffirm the main claim of this paper, that $\mathcal{S}<0$ must hold.

\subsection{Adiabatic sound speed}
Assuming for a moment that the pressure $p_{\rm DE}$ depends on the entropy $S$ and energy density $\rho_{\rm DE}$, a generic variation can be written as $\delta p_{\rm DE}(S,\rho_{\rm DE})=(\partial p/\partial S)\delta S + (\partial p/\partial \rho_{\rm DE})\delta \rho_{\rm DE}$. We can rewrite this as $\delta p=\delta p_{\rm na} + c_a^2\delta\rho_{\rm DE}$, where $\delta p_{\rm na}$ is the non-adiabatic perturbation related to a variation in the entropy $S$, and $c_a^2$ is the adiabatic sound speed, which can be written as
\begin{equation}
    c_a^2 = \frac{\dot{p}_{\rm DE}}{\dot{\rho}_{\rm DE}}.
\end{equation}
This definition follows naturally by considering the behaviour of $\delta p_{\rm DE}$ and $\delta \rho_{\rm DE}$ under the gauge transformation $t\to t-\delta t$, $\delta\rho_{\rm DE}\to\dot{\rho}\delta t$, $\delta p_{\rm DE}\to\delta p_{\rm DE}+\dot{p}_{\rm DE}\delta t$, where only the definition of $c_a^2$ leaves $\delta p_{\rm na}$ gauge invariant \cite{Christopherson:2008ry,Malik:2008yp}. 

\begin{figure}[h]
    \centering
    \includegraphics[scale=0.8]{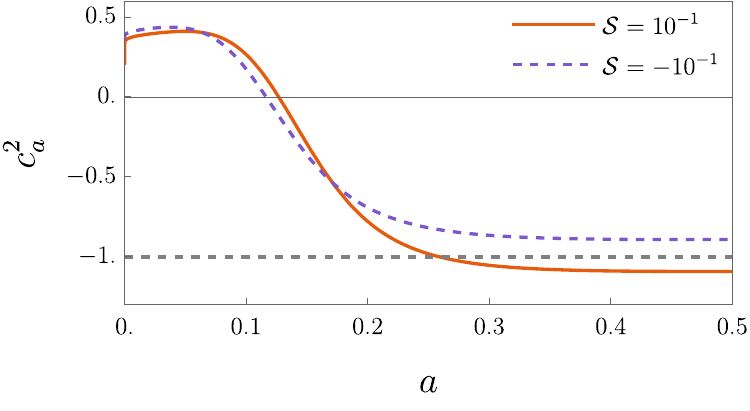}
    \caption{The adiabatic sound speed for different values of the coefficient $\mathcal{S}$.}
    \label{fig:ca2}
\end{figure}

Since the adiabatic sound speed can be written using only background quantities, we can find it without resorting to perturbation theory; the result is rather lengthy and can be found in \ref{app:longs}. We find that for small values of the scale factor, $c_a^2\to1/3$, and then increases slightly before relaxing smoothly down to a value close to minus unity for larger values of $a$, as can be seen in Figure~\ref{fig:ca2}.
This behaviour is similar to that of IR-modified Ho\v{r}ava-Lifshitz gravity, which has also been shown to contain a type of dynamical dark energy, but where $c_a^2$ flows from $+1/3 \to -1/3$ \cite{Park:2009zr,DiValentino:2022eot}. We note that that a negative adiabatic sound speed is not a problem, as it does not describe the propagation speed, but rather the relative change between the pressure and the density.

\subsection{Null-energy condition}
\begin{figure}[h]
    \centering
    \includegraphics[scale=0.8]{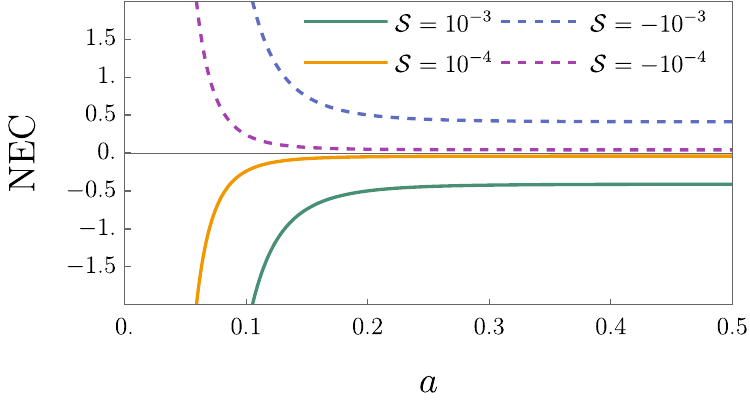}
    \caption{The Null Energy Condition (NEC) in arbitrary units, which can be seen to be violated for positive values of the coefficient $\mathcal{S}$.}
    \label{fig:nec}
\end{figure}

The Null-Energy Condition (NEC) plays an important role in general relativity, where it is an ingredient in the Hawking-Penrose singularity theorem and the positive mass theorem. For a causal and Lorentz-invariant scalar-field theory, imposing the NEC is sufficient to guarantee stability, and NEC violation is often used as an indication of phantom behaviour. It states that for any null vector $k^\mu$, the stress-energy tensor should satisfy
\begin{equation}
    T_{\mu\nu}k^\mu k^\nu \geq0,
\end{equation}
and can be interpreted as a condition of causality in the theory; the equivalent condition reads $\rho+p\geq0$. Since we are working with with a model where local Lorentz invariance is broken explicitly, there is no guarantee that the dynamical dark energy discussed here will uphold causality, and therefore the NEC may be in jeopardy. We check this explicitly by plotting $\rho_{\rm DE}+p_{\rm DE}$ for different values of the coefficient $\mathcal{S}$, which can be seen in Figure~\ref{fig:nec}, after which it becomes clear that $\mathcal{S}$ {\it needs to be negative} for the NEC to hold.
We can come to the same conclusion by studying the modified continuity equation (\ref{eq:cont}), where the NEC implies that energy density cannot increase as long as the Universe is expanding. In order for the NEC to hold here, the auxiliary function $f(\mathcal{S},w)$ must have the correct sign, which can easily be shown to occur only for $\mathcal{S}<0$ when $w=-1$ and for all values of $\mathcal{S}$ when $w=1/3$. 


\section{Conclusions}\label{sec:disc}
In this paper, we have investigated the DE properties of a cosmological solution featuring a simple type of spacetime-symmetry breaking. By imposing the covariant conservation of the entire stress-energy tensor (without demanding that the matter stress-energy tensor be conserved separately) we endow the model with a modified continuity equation which affects the evolution of the cosmological constant and the radiation density; it would be interesting to identify if other types of spacetime-symmetry breaking could isolate the modification to one constituent only.

By writing down the modified Friedmann equations and considered the non-$\Lambda$CDM contributions as a type of dynamical DE, we identified the effective DE equation of state $w_{\rm DE}$. We found that $w_{\rm DE}$ is singular at small values of the scale factor $a$ when the spacetime-symmetry breaking coefficient $\mathcal{S}$ is positive, but exhibits smooth evolution for negative values of the same; we also found that the DE is phantom for large $a$ if $\mathcal{S}>0$. Further, we identified the CPL parameters of our DE model and concluded that current bounds on $w_0$ and $w_a$ yield weaker constraints constraints on the coefficient for spacetime-symmetry breaking $\mathcal{S}$ as compared to other results. It would be interesting to implement this model into a Markov-Chain Monte Carlo (MCMC) code and fully exploit the quadratic dependence of $\mathcal{S}$ in the CPL parameters; we leave this for future work. We also investigated the adiabatic sound speed, which shows no discontinuities for any value of $\mathcal{S}$, but we concluded that the NEC is broken for $\mathcal{S}>0$. We also compare with a logarithmic parametrisation of the barotropic index, and obtain a poor fit; however, we do find that the parameters obtain the correct signs when $\mathcal{S}<0$.

The main take-home message from this analysis is that in this specific realisation of explicit breaking, the EFT {\it coefficient $\mathcal{S}$ needs to be negative} (or zero) if one wants to avoid the issues normally present in phantom-like cosmological fluids.
A negative $\mathcal{S}$ would have consequences in other contexts: for example, $\mathcal{S}$ is related to the propagation speed of gravitational waves, as was discussed in \cite{Nilsson:2022mzq}, where the bound $-6\cdot 10^{-15}<\mathcal{S}<+7\cdot 10^{-16}$ was obtained from the observation of GW170817 and GRB170817A, and restricting $\mathcal{S}$ to negative values implies that the propagation speed of tensor modes is less than unity, i.e. {\it slower} than light; generalising the initial ansatz (\ref{eq:smunu}) will necessarily alter these predictions. We also expect severe constraints from early-Universe physics such as BBN, which will be affected given that the logarithmic contribution to the equation of state will dominate at $a < 10^{-3}$. In this era, this contribution may be thought of as a time-dependent effective number of relativistic species ($N_{\rm eff}$) and may also affect the radius of the sound horizon, although the above constraint from the speed of gravity makes any modification necessarily very small. Comparison with early-Universe data is planned, but lies beyond the scope of the present work.

\section*{Acknowledgements}
The author was financed by CNES and IBS under the project code IBS-R018-D3. The authors acknowledges support by PSL/Observatoire de Paris and thanks Eoin \'{O} Colg\'{a}in and Quentin G. Bailey for useful comments.

\appendix

\section{Explicit expressions for $w_{\rm DE}$ and $c_a^2$}\label{app:longs}
The expressions for the equation of state and adiabatic sound speed are
\tiny
\begin{equation}\label{eq:wDE}
    \begin{aligned}
        w_{\rm DE} &=\frac{-3 a^4 (\mathcal{S} (5 \mathcal{S}+2)+2) \Omega_\Lambda^0 +\mathcal{S} \ln (a) \left(-9 a^4 (7 \mathcal{S}+2) \Omega_\Lambda^0 +\mathcal{S} \ln (a) \left(\Omega_r^0-27 a^4 \Omega_\Lambda^0 \right)-(\mathcal{S}-2) \Omega_r^0\right)-\mathcal{S} (\mathcal{S}+2) \Omega_r^0}{3 \mathcal{S} \ln
   (a) \left(3 a^4 (5 \mathcal{S}+2) \Omega_\Lambda^0 +\mathcal{S} \ln (a) \left(9 a^4 \Omega_\Lambda^0 +\Omega_r^0\right)+(\mathcal{S}+2) \Omega_r^0\right)+6 a^4 \Omega_\Lambda^0 },\\[5mm]
    c_a^2 &=\frac{1}{3}-\frac{2 \left(-39 a^4 \mathcal{S} \Omega_\Lambda^0 -36 a^4 \mathcal{S} \Omega_\Lambda^0  \ln (a)-12 a^4 \Omega_\Lambda^0 +4 \mathcal{S} \Omega_r^0 \ln (a)+\mathcal{S} \Omega_r^0+4 \Omega_r^0\right)}{3 \left(-15 a^4 \mathcal{S} \Omega_\Lambda^0 -18 a^4 \mathcal{S} \Omega_\Lambda^0  \ln (a)-6 a^4 \Omega_\Lambda^0 +4 \mathcal{S} \Omega_r^0 \ln ^2(a)+2 \mathcal{S} \Omega_r^0 \ln (a)+8 \Omega_r^0 \ln (a)-\mathcal{S} \Omega_r^0-2 \Omega_r^0\right)}.
    \end{aligned}
\end{equation}
\normalsize

\section{Sanity check of modified conservation laws}\label{app:sanity}
In this appendix, we show explicitly that the modified continuity equation (\ref{eq:cont}) implies the effective conservation laws for the perfect-fluid constituents (\ref{eq:omegamods}). 

We start by assuming that, as in GR, there is no mixing between the different constituents. By using the generic perfect-fluid stress-energy tensor in comoving coordinates
\begin{equation}
    (T_M)^{\mu\nu} = \text{diag}(\sum_l\rho_l~, \sum_l(p_l,p_l,p_l)),
\end{equation}
and so we can write the total modified Bianchi identity in Eq.~(\ref{eq:bianchi0}) as
\begin{equation}\label{eq:B2}
    6\mathcal{S}\frac{\dot{a}}{a}\frac{\ddot{a}}{a}+3\mathcal{S}\frac{\dddot{a}}{a} -\sum_i(\dot{\rho}_i+3H(\rho_i+p_i)) = 0.
\end{equation}
In order to evaluate the left-hand side, we use the Friedmann equations (\ref{eq:fried2}) for $k=0$, which we can write as
\begin{equation}
    \begin{aligned}
        \left(\frac{\dot{a}}{a}\right)^2 &= \sum_i\left[\frac{\kappa \rho_i}{3(1-\tfrac{3}{2}\mathcal{S})}+\frac{\kappa p_i \mathcal{S}}{(2-3\mathcal{S})(1-\mathcal{S})}\right] \\
        \frac{\ddot{a}}{a} &= -\sum_i\left[\frac{\rho_i+3p_i}{6(1-\tfrac{3}{2}\mathcal{S})}\right],
    \end{aligned}
\end{equation}
and thus Eq.~(\ref{eq:B2}) can be manipulated into the form
\begin{equation}
    \sum_i\left[-6H\rho_i(1-\mathcal{S}+w)-\dot{\rho}_i(2(1-\mathcal{S})+3w\mathcal{S})\right] = 0.
\end{equation}
Since we have no mixing between the perfect-fluid constituents, the following must therefore hold for all $i$
\begin{equation}
    \dot{\rho}_i+3H\underbrace{\left(\frac{2(1-\mathcal{S}+w)}{2(1-\mathcal{S})+3w\mathcal{S}}\right)}_{\equiv f(w,\mathcal{S})}\rho_i = 0
\end{equation}

\bibliographystyle{elsarticle-num} 
\bibliography{example}

\providecommand{\noopsort}[1]{}\providecommand{\singleletter}[1]{#1}%
\begin{thebibliography}{10}
\expandafter\ifx\csname url\endcsname\relax
  \def\url#1{\texttt{#1}}\fi
\expandafter\ifx\csname urlprefix\endcsname\relax\def\urlprefix{URL }\fi
\expandafter\ifx\csname href\endcsname\relax
  \def\href#1#2{#2} \def\path#1{#1}\fi

\bibitem{SupernovaSearchTeam:1998fmf}
A.~G. Riess, et~al., {Observational evidence from supernovae for an
  accelerating universe and a cosmological constant}, Astron. J. 116 (1998)
  1009--1038.
\newblock \href {http://arxiv.org/abs/astro-ph/9805201}
  {\path{arXiv:astro-ph/9805201}}, \href {https://doi.org/10.1086/300499}
  {\path{doi:10.1086/300499}}.

\bibitem{SupernovaCosmologyProject:1998vns}
S.~Perlmutter, et~al., {Measurements of $\Omega$ and $\Lambda$ from 42 high
  redshift supernovae}, Astrophys. J. 517 (1999) 565--586.
\newblock \href {http://arxiv.org/abs/astro-ph/9812133}
  {\path{arXiv:astro-ph/9812133}}, \href {https://doi.org/10.1086/307221}
  {\path{doi:10.1086/307221}}.

\bibitem{Planck:2018vyg}
N.~Aghanim, et~al., {Planck 2018 results. VI. Cosmological parameters}, Astron.
  Astrophys. 641 (2020) A6, [Erratum: Astron.Astrophys. 652, C4 (2021)].
\newblock \href {http://arxiv.org/abs/1807.06209} {\path{arXiv:1807.06209}},
  \href {https://doi.org/10.1051/0004-6361/201833910}
  {\path{doi:10.1051/0004-6361/201833910}}.

\bibitem{Martin:2012bt}
J.~Martin, {Everything You Always Wanted To Know About The Cosmological
  Constant Problem (But Were Afraid To Ask)}, Comptes Rendus Physique 13 (2012)
  566--665.
\newblock \href {http://arxiv.org/abs/1205.3365} {\path{arXiv:1205.3365}},
  \href {https://doi.org/10.1016/j.crhy.2012.04.008}
  {\path{doi:10.1016/j.crhy.2012.04.008}}.

\bibitem{Zlatev:1998tr}
I.~Zlatev, L.-M. Wang, P.~J. Steinhardt, {Quintessence, cosmic coincidence, and
  the cosmological constant}, Phys. Rev. Lett. 82 (1999) 896--899.
\newblock \href {http://arxiv.org/abs/astro-ph/9807002}
  {\path{arXiv:astro-ph/9807002}}, \href
  {https://doi.org/10.1103/PhysRevLett.82.896}
  {\path{doi:10.1103/PhysRevLett.82.896}}.

\bibitem{Keeley:2022ojz}
R.~E. Keeley, A.~Shafieloo, {Ruling Out New Physics at Low Redshift as a
  Solution to the H0 Tension}, Phys. Rev. Lett. 131~(11) (2023) 111002.
\newblock \href {http://arxiv.org/abs/2206.08440} {\path{arXiv:2206.08440}},
  \href {https://doi.org/10.1103/PhysRevLett.131.111002}
  {\path{doi:10.1103/PhysRevLett.131.111002}}.

\bibitem{Krishnan:2021dyb}
C.~Krishnan, R.~Mohayaee, E.~O. Colg\'ain, M.~M. Sheikh-Jabbari, L.~Yin, {Does
  Hubble tension signal a breakdown in FLRW cosmology?}, Class. Quant. Grav.
  38~(18) (2021) 184001.
\newblock \href {http://arxiv.org/abs/2105.09790} {\path{arXiv:2105.09790}},
  \href {https://doi.org/10.1088/1361-6382/ac1a81}
  {\path{doi:10.1088/1361-6382/ac1a81}}.

\bibitem{Ratra:1987rm}
B.~Ratra, P.~J.~E. Peebles, {Cosmological Consequences of a Rolling Homogeneous
  Scalar Field}, Phys. Rev. D 37 (1988) 3406.
\newblock \href {https://doi.org/10.1103/PhysRevD.37.3406}
  {\path{doi:10.1103/PhysRevD.37.3406}}.

\bibitem{Wetterich:1987fm}
C.~Wetterich, {Cosmology and the Fate of Dilatation Symmetry}, Nucl. Phys. B
  302 (1988) 668--696.
\newblock \href {http://arxiv.org/abs/1711.03844} {\path{arXiv:1711.03844}},
  \href {https://doi.org/10.1016/0550-3213(88)90193-9}
  {\path{doi:10.1016/0550-3213(88)90193-9}}.

\bibitem{Armendariz-Picon:1999hyi}
C.~Armendariz-Picon, T.~Damour, V.~F. Mukhanov, {k - inflation}, Phys. Lett. B
  458 (1999) 209--218.
\newblock \href {http://arxiv.org/abs/hep-th/9904075}
  {\path{arXiv:hep-th/9904075}}, \href
  {https://doi.org/10.1016/S0370-2693(99)00603-6}
  {\path{doi:10.1016/S0370-2693(99)00603-6}}.

\bibitem{Chiba:1999ka}
T.~Chiba, T.~Okabe, M.~Yamaguchi, {Kinetically driven quintessence}, Phys. Rev.
  D 62 (2000) 023511.
\newblock \href {http://arxiv.org/abs/astro-ph/9912463}
  {\path{arXiv:astro-ph/9912463}}, \href
  {https://doi.org/10.1103/PhysRevD.62.023511}
  {\path{doi:10.1103/PhysRevD.62.023511}}.

\bibitem{Frusciante:2019xia}
N.~Frusciante, L.~Perenon, {Effective field theory of dark energy: A review},
  Phys. Rept. 857 (2020) 1--63.
\newblock \href {http://arxiv.org/abs/1907.03150} {\path{arXiv:1907.03150}},
  \href {https://doi.org/10.1016/j.physrep.2020.02.004}
  {\path{doi:10.1016/j.physrep.2020.02.004}}.

\bibitem{Ludwick:2017tox}
K.~J. Ludwick, {The viability of phantom dark energy: A review}, Mod. Phys.
  Lett. A 32~(28) (2017) 1730025.
\newblock \href {http://arxiv.org/abs/1708.06981} {\path{arXiv:1708.06981}},
  \href {https://doi.org/10.1142/S0217732317300257}
  {\path{doi:10.1142/S0217732317300257}}.

\bibitem{Dabrowski:2003jm}
M.~P. Dabrowski, T.~Stachowiak, M.~Szydlowski, {Phantom cosmologies}, Phys.
  Rev. D 68 (2003) 103519.
\newblock \href {http://arxiv.org/abs/hep-th/0307128}
  {\path{arXiv:hep-th/0307128}}, \href
  {https://doi.org/10.1103/PhysRevD.68.103519}
  {\path{doi:10.1103/PhysRevD.68.103519}}.

\bibitem{Nojiri:2005sx}
S.~Nojiri, S.~D. Odintsov, S.~Tsujikawa, {Properties of singularities in
  (phantom) dark energy universe}, Phys. Rev. D 71 (2005) 063004.
\newblock \href {http://arxiv.org/abs/hep-th/0501025}
  {\path{arXiv:hep-th/0501025}}, \href
  {https://doi.org/10.1103/PhysRevD.71.063004}
  {\path{doi:10.1103/PhysRevD.71.063004}}.

\bibitem{Pollock:1988xe}
M.~D. Pollock, {On the Initial Conditions for Superexponential Inflation},
  Phys. Lett. B 215 (1988) 635--641.
\newblock \href {https://doi.org/10.1016/0370-2693(88)90034-2}
  {\path{doi:10.1016/0370-2693(88)90034-2}}.

\bibitem{Torres:2002pe}
D.~F. Torres, {Quintessence, superquintessence and observable quantities in
  Brans-Dicke and nonminimally coupled theories}, Phys. Rev. D 66 (2002)
  043522.
\newblock \href {http://arxiv.org/abs/astro-ph/0204504}
  {\path{arXiv:astro-ph/0204504}}, \href
  {https://doi.org/10.1103/PhysRevD.66.043522}
  {\path{doi:10.1103/PhysRevD.66.043522}}.

\bibitem{Buniy:2006xf}
R.~V. Buniy, S.~D.~H. Hsu, B.~M. Murray, {The Null energy condition and
  instability}, Phys. Rev. D 74 (2006) 063518.
\newblock \href {http://arxiv.org/abs/hep-th/0606091}
  {\path{arXiv:hep-th/0606091}}, \href
  {https://doi.org/10.1103/PhysRevD.74.063518}
  {\path{doi:10.1103/PhysRevD.74.063518}}.

\bibitem{Dubovsky:2005xd}
S.~Dubovsky, T.~Gregoire, A.~Nicolis, R.~Rattazzi, {Null energy condition and
  superluminal propagation}, JHEP 03 (2006) 025.
\newblock \href {http://arxiv.org/abs/hep-th/0512260}
  {\path{arXiv:hep-th/0512260}}, \href
  {https://doi.org/10.1088/1126-6708/2006/03/025}
  {\path{doi:10.1088/1126-6708/2006/03/025}}.

\bibitem{Arkani-Hamed:2003pdi}
N.~Arkani-Hamed, H.-C. Cheng, M.~A. Luty, S.~Mukohyama, {Ghost condensation and
  a consistent infrared modification of gravity}, JHEP 05 (2004) 074.
\newblock \href {http://arxiv.org/abs/hep-th/0312099}
  {\path{arXiv:hep-th/0312099}}, \href
  {https://doi.org/10.1088/1126-6708/2004/05/074}
  {\path{doi:10.1088/1126-6708/2004/05/074}}.

\bibitem{Nicolis:2008in}
A.~Nicolis, R.~Rattazzi, E.~Trincherini, {The Galileon as a local modification
  of gravity}, Phys. Rev. D 79 (2009) 064036.
\newblock \href {http://arxiv.org/abs/0811.2197} {\path{arXiv:0811.2197}},
  \href {https://doi.org/10.1103/PhysRevD.79.064036}
  {\path{doi:10.1103/PhysRevD.79.064036}}.

\bibitem{Rubakov:2006pn}
V.~A. Rubakov, {Phantom without UV pathology}, Theor. Math. Phys. 149 (2006)
  1651--1664.
\newblock \href {http://arxiv.org/abs/hep-th/0604153}
  {\path{arXiv:hep-th/0604153}}, \href
  {https://doi.org/10.1007/s11232-006-0149-7}
  {\path{doi:10.1007/s11232-006-0149-7}}.

\bibitem{Banerjee:2020xcn}
A.~Banerjee, H.~Cai, L.~Heisenberg, E.~O. Colg\'ain, M.~M. Sheikh-Jabbari,
  T.~Yang, {Hubble sinks in the low-redshift swampland}, Phys. Rev. D 103~(8)
  (2021) L081305.
\newblock \href {http://arxiv.org/abs/2006.00244} {\path{arXiv:2006.00244}},
  \href {https://doi.org/10.1103/PhysRevD.103.L081305}
  {\path{doi:10.1103/PhysRevD.103.L081305}}.

\bibitem{Lee:2022cyh}
B.-H. Lee, W.~Lee, E.~O. Colg\'ain, M.~M. Sheikh-Jabbari, S.~Thakur, {Is local
  H $_{0}$ at odds with dark energy EFT?}, JCAP 04~(04) (2022) 004.
\newblock \href {http://arxiv.org/abs/2202.03906} {\path{arXiv:2202.03906}},
  \href {https://doi.org/10.1088/1475-7516/2022/04/004}
  {\path{doi:10.1088/1475-7516/2022/04/004}}.

\bibitem{DiValentino:2021izs}
E.~Di~Valentino, O.~Mena, S.~Pan, L.~Visinelli, W.~Yang, A.~Melchiorri, D.~F.
  Mota, A.~G. Riess, J.~Silk, {In the realm of the Hubble tension\textemdash{}a
  review of solutions}, Class. Quant. Grav. 38~(15) (2021) 153001.
\newblock \href {http://arxiv.org/abs/2103.01183} {\path{arXiv:2103.01183}},
  \href {https://doi.org/10.1088/1361-6382/ac086d}
  {\path{doi:10.1088/1361-6382/ac086d}}.

\bibitem{Riess:2020fzl}
A.~G. Riess, S.~Casertano, W.~Yuan, J.~B. Bowers, L.~Macri, J.~C. Zinn,
  D.~Scolnic, {Cosmic Distances Calibrated to 1\% Precision with Gaia EDR3
  Parallaxes and Hubble Space Telescope Photometry of 75 Milky Way Cepheids
  Confirm Tension with $\Lambda$CDM}, Astrophys. J. Lett. 908~(1) (2021) L6.
\newblock \href {http://arxiv.org/abs/2012.08534} {\path{arXiv:2012.08534}},
  \href {https://doi.org/10.3847/2041-8213/abdbaf}
  {\path{doi:10.3847/2041-8213/abdbaf}}.

\bibitem{Freedman:2021ahq}
W.~L. Freedman, {Measurements of the Hubble Constant: Tensions in Perspective},
  Astrophys. J. 919~(1) (2021) 16.
\newblock \href {http://arxiv.org/abs/2106.15656} {\path{arXiv:2106.15656}},
  \href {https://doi.org/10.3847/1538-4357/ac0e95}
  {\path{doi:10.3847/1538-4357/ac0e95}}.

\bibitem{Blas:2011en}
D.~Blas, S.~Sibiryakov, {Technically natural dark energy from Lorentz
  breaking}, JCAP 07 (2011) 026.
\newblock \href {http://arxiv.org/abs/1104.3579} {\path{arXiv:1104.3579}},
  \href {https://doi.org/10.1088/1475-7516/2011/07/026}
  {\path{doi:10.1088/1475-7516/2011/07/026}}.

\bibitem{Audren:2013dwa}
B.~Audren, D.~Blas, J.~Lesgourgues, S.~Sibiryakov, {Cosmological constraints on
  Lorentz violating dark energy}, JCAP 08 (2013) 039.
\newblock \href {http://arxiv.org/abs/1305.0009} {\path{arXiv:1305.0009}},
  \href {https://doi.org/10.1088/1475-7516/2013/08/039}
  {\path{doi:10.1088/1475-7516/2013/08/039}}.

\bibitem{Park:2009zr}
M.-i. Park, {A Test of Horava Gravity: The Dark Energy}, JCAP 01 (2010) 001.
\newblock \href {http://arxiv.org/abs/0906.4275} {\path{arXiv:0906.4275}},
  \href {https://doi.org/10.1088/1475-7516/2010/01/001}
  {\path{doi:10.1088/1475-7516/2010/01/001}}.

\bibitem{Saridakis:2009bv}
E.~N. Saridakis, {Horava-Lifshitz Dark Energy}, Eur. Phys. J. C 67 (2010)
  229--235.
\newblock \href {http://arxiv.org/abs/0905.3532} {\path{arXiv:0905.3532}},
  \href {https://doi.org/10.1140/epjc/s10052-010-1294-6}
  {\path{doi:10.1140/epjc/s10052-010-1294-6}}.

\bibitem{DiValentino:2022eot}
E.~Di~Valentino, N.~A. Nilsson, M.-I. Park, {A new test of dynamical dark
  energy models and cosmic tensions in Ho\v{r}ava gravity}, Mon. Not. Roy.
  Astron. Soc. 519~(4) (2023) 5043--5058.
\newblock \href {http://arxiv.org/abs/2212.07683} {\path{arXiv:2212.07683}},
  \href {https://doi.org/10.1093/mnras/stac3824}
  {\path{doi:10.1093/mnras/stac3824}}.

\bibitem{ONeal-Ault:2020ebv}
K.~O'Neal-Ault, Q.~G. Bailey, N.~A. Nilsson, {3+1 formulation of the standard
  model extension gravity sector}, Phys. Rev. D 103~(4) (2021) 044010.
\newblock \href {http://arxiv.org/abs/2009.00949} {\path{arXiv:2009.00949}},
  \href {https://doi.org/10.1103/PhysRevD.103.044010}
  {\path{doi:10.1103/PhysRevD.103.044010}}.

\bibitem{Colladay:1996iz}
D.~Colladay, V.~A. Kostelecky, {CPT violation and the standard model}, Phys.
  Rev. D 55 (1997) 6760--6774.
\newblock \href {http://arxiv.org/abs/hep-ph/9703464}
  {\path{arXiv:hep-ph/9703464}}, \href
  {https://doi.org/10.1103/PhysRevD.55.6760}
  {\path{doi:10.1103/PhysRevD.55.6760}}.

\bibitem{Colladay:1998fq}
D.~Colladay, V.~A. Kostelecky, {Lorentz violating extension of the standard
  model}, Phys. Rev. D 58 (1998) 116002.
\newblock \href {http://arxiv.org/abs/hep-ph/9809521}
  {\path{arXiv:hep-ph/9809521}}, \href
  {https://doi.org/10.1103/PhysRevD.58.116002}
  {\path{doi:10.1103/PhysRevD.58.116002}}.

\bibitem{Kostelecky:2003fs}
V.~A. Kostelecky, {Gravity, Lorentz violation, and the standard model}, Phys.
  Rev. D 69 (2004) 105009.
\newblock \href {http://arxiv.org/abs/hep-th/0312310}
  {\path{arXiv:hep-th/0312310}}, \href
  {https://doi.org/10.1103/PhysRevD.69.105009}
  {\path{doi:10.1103/PhysRevD.69.105009}}.

\bibitem{Kostelecky:2008ts}
V.~A. Kostelecky, N.~Russell, {Data Tables for Lorentz and CPT Violation} (1
  2008).
\newblock \href {http://arxiv.org/abs/0801.0287} {\path{arXiv:0801.0287}}.

\bibitem{Bonder:2017dpb}
Y.~Bonder, G.~Leon, {Inflation as an amplifier: the case of Lorentz violation},
  Phys. Rev. D 96~(4) (2017) 044036.
\newblock \href {http://arxiv.org/abs/1704.05894} {\path{arXiv:1704.05894}},
  \href {https://doi.org/10.1103/PhysRevD.96.044036}
  {\path{doi:10.1103/PhysRevD.96.044036}}.

\bibitem{Nilsson:2022mzq}
N.~A. Nilsson, {Explicit spacetime-symmetry breaking and the dynamics of
  primordial fields}, Phys. Rev. D 106~(10) (2022) 104036.
\newblock \href {http://arxiv.org/abs/2205.00496} {\path{arXiv:2205.00496}},
  \href {https://doi.org/10.1103/PhysRevD.106.104036}
  {\path{doi:10.1103/PhysRevD.106.104036}}.

\bibitem{Reyes:2022dil}
C.~M. Reyes, M.~Schreck, A.~Soto, {Cosmology in the presence of
  diffeomorphism-violating, nondynamical background fields}, Phys. Rev. D
  106~(2) (2022) 023524.
\newblock \href {http://arxiv.org/abs/2205.06329} {\path{arXiv:2205.06329}},
  \href {https://doi.org/10.1103/PhysRevD.106.023524}
  {\path{doi:10.1103/PhysRevD.106.023524}}.

\bibitem{Khodadi:2023ezj}
M.~Khodadi, M.~Schreck, {Hubble tension as a guide for refining the early
  Universe: Cosmologies with explicit local Lorentz and diffeomorphism
  violation}, Phys. Dark Univ. 39 (2023) 101170.
\newblock \href {http://arxiv.org/abs/2301.03883} {\path{arXiv:2301.03883}},
  \href {https://doi.org/10.1016/j.dark.2023.101170}
  {\path{doi:10.1016/j.dark.2023.101170}}.

\bibitem{Maluf:2021lwh}
R.~V. Maluf, J.~C.~S. Neves, {Bumblebee field as a source of cosmological
  anisotropies}, JCAP 10 (2021) 038.
\newblock \href {http://arxiv.org/abs/2105.08659} {\path{arXiv:2105.08659}},
  \href {https://doi.org/10.1088/1475-7516/2021/10/038}
  {\path{doi:10.1088/1475-7516/2021/10/038}}.

\bibitem{Bailey:2006fd}
Q.~G. Bailey, V.~A. Kostelecky, {Signals for Lorentz violation in
  post-Newtonian gravity}, Phys. Rev. D 74 (2006) 045001.
\newblock \href {http://arxiv.org/abs/gr-qc/0603030}
  {\path{arXiv:gr-qc/0603030}}, \href
  {https://doi.org/10.1103/PhysRevD.74.045001}
  {\path{doi:10.1103/PhysRevD.74.045001}}.

\bibitem{Hees:2013iv}
A.~Hees, B.~Lamine, S.~Reynaud, M.~T. Jaekel, C.~Le~Poncin-Lafitte, V.~Lainey,
  A.~Fuzfa, J.~M. Courty, V.~Dehant, P.~Wolf, {Simulations of Solar System
  observations in alternative theories of gravity}, in: {13th Marcel Grossmann
  Meeting on Recent Developments in Theoretical and Experimental General
  Relativity, Astrophysics, and Relativistic Field Theories}, 2015, pp.
  2357--2359.
\newblock \href {http://arxiv.org/abs/1301.1658} {\path{arXiv:1301.1658}},
  \href {https://doi.org/10.1142/9789814623995_0440}
  {\path{doi:10.1142/9789814623995_0440}}.

\bibitem{LePoncin-Lafitte:2016ocy}
C.~Le~Poncin-Lafitte, A.~Hees, S.~Lambert, {Lorentz symmetry and Very Long
  Baseline Interferometry}, Phys. Rev. D 94~(12) (2016) 125030.
\newblock \href {http://arxiv.org/abs/1604.01663} {\path{arXiv:1604.01663}},
  \href {https://doi.org/10.1103/PhysRevD.94.125030}
  {\path{doi:10.1103/PhysRevD.94.125030}}.

\bibitem{Long:2014swa}
J.~C. Long, V.~A. Kosteleck\'y, {Search for Lorentz violation in short-range
  gravity}, Phys. Rev. D 91~(9) (2015) 092003.
\newblock \href {http://arxiv.org/abs/1412.8362} {\path{arXiv:1412.8362}},
  \href {https://doi.org/10.1103/PhysRevD.91.092003}
  {\path{doi:10.1103/PhysRevD.91.092003}}.

\bibitem{Shao:2016cjk}
C.-G. Shao, et~al., {Combined search for Lorentz violation in short-range
  gravity}, Phys. Rev. Lett. 117~(7) (2016) 071102.
\newblock \href {http://arxiv.org/abs/1607.06095} {\path{arXiv:1607.06095}},
  \href {https://doi.org/10.1103/PhysRevLett.117.071102}
  {\path{doi:10.1103/PhysRevLett.117.071102}}.

\bibitem{Shao:2018lsx}
C.-G. Shao, Y.-F. Chen, Y.-J. Tan, S.-Q. Yang, J.~Luo, M.~E. Tobar, J.~C. Long,
  E.~Weisman, V.~A. Kosteleck\'y, {Combined Search for a Lorentz-Violating
  Force in Short-Range Gravity Varying as the Inverse Sixth Power of Distance},
  Phys. Rev. Lett. 122~(1) (2019) 011102.
\newblock \href {http://arxiv.org/abs/1812.11123} {\path{arXiv:1812.11123}},
  \href {https://doi.org/10.1103/PhysRevLett.122.011102}
  {\path{doi:10.1103/PhysRevLett.122.011102}}.

\bibitem{Bailey:2022wuv}
Q.~G. Bailey, J.~L. James, J.~R. Slone, K.~O'Neal-Ault, {Short-range forces due
  to Lorentz-symmetry violation}, Class. Quant. Grav. 40~(4) (2023) 045006.
\newblock \href {http://arxiv.org/abs/2210.00605} {\path{arXiv:2210.00605}},
  \href {https://doi.org/10.1088/1361-6382/acb0ab}
  {\path{doi:10.1088/1361-6382/acb0ab}}.

\bibitem{Shao:2018vul}
L.~Shao, Q.~G. Bailey, {Testing velocity-dependent CPT-violating gravitational
  forces with radio pulsars}, Phys. Rev. D 98~(8) (2018) 084049.
\newblock \href {http://arxiv.org/abs/1810.06332} {\path{arXiv:1810.06332}},
  \href {https://doi.org/10.1103/PhysRevD.98.084049}
  {\path{doi:10.1103/PhysRevD.98.084049}}.

\bibitem{Shao:2014oha}
L.~Shao, {Tests of local Lorentz invariance violation of gravity in the
  standard model extension with pulsars}, Phys. Rev. Lett. 112 (2014) 111103.
\newblock \href {http://arxiv.org/abs/1402.6452} {\path{arXiv:1402.6452}},
  \href {https://doi.org/10.1103/PhysRevLett.112.111103}
  {\path{doi:10.1103/PhysRevLett.112.111103}}.

\bibitem{ONeal-Ault:2021uwu}
K.~O'Neal-Ault, Q.~G. Bailey, T.~Dumerchat, L.~Haegel, J.~Tasson, {Analysis of
  Birefringence and Dispersion Effects from Spacetime-Symmetry Breaking in
  Gravitational Waves}, Universe 7~(10) (2021) 380.
\newblock \href {http://arxiv.org/abs/2108.06298} {\path{arXiv:2108.06298}},
  \href {https://doi.org/10.3390/universe7100380}
  {\path{doi:10.3390/universe7100380}}.

\bibitem{Wang:2021ctl}
Z.~Wang, L.~Shao, C.~Liu, {New Limits on the Lorentz/CPT Symmetry Through 50
  Gravitational-wave Events}, Astrophys. J. 921~(2) (2021) 158.
\newblock \href {http://arxiv.org/abs/2108.02974} {\path{arXiv:2108.02974}},
  \href {https://doi.org/10.3847/1538-4357/ac223c}
  {\path{doi:10.3847/1538-4357/ac223c}}.

\bibitem{Liu:2020slm}
X.~Liu, V.~F. He, T.~M. Mikulski, D.~Palenova, C.~E. Williams, J.~Creighton,
  J.~D. Tasson, {Measuring the speed of gravitational waves from the first and
  second observing run of Advanced LIGO and Advanced Virgo}, Phys. Rev. D
  102~(2) (2020) 024028.
\newblock \href {http://arxiv.org/abs/2005.03121} {\path{arXiv:2005.03121}},
  \href {https://doi.org/10.1103/PhysRevD.102.024028}
  {\path{doi:10.1103/PhysRevD.102.024028}}.

\bibitem{Bisnovatyi-Kogan:2023frj}
G.~S. Bisnovatyi-Kogan, A.~M. Nikishin, {Eliminating the Hubble Tension in the
  Presence of the Interconnection between Dark Energy and Matter in the Modern
  Universe}, Astron. Rep. 67~(2) (2023) 115--124.
\newblock \href {http://arxiv.org/abs/2305.17722} {\path{arXiv:2305.17722}},
  \href {https://doi.org/10.1134/S1063772923020038}
  {\path{doi:10.1134/S1063772923020038}}.

\bibitem{Carroll:2004ai}
S.~M. Carroll, E.~A. Lim, {Lorentz-violating vector fields slow the universe
  down}, Phys. Rev. D 70 (2004) 123525.
\newblock \href {http://arxiv.org/abs/hep-th/0407149}
  {\path{arXiv:hep-th/0407149}}, \href
  {https://doi.org/10.1103/PhysRevD.70.123525}
  {\path{doi:10.1103/PhysRevD.70.123525}}.

\bibitem{LIGOScientific:2017zic}
B.~P. Abbott, et~al., {Gravitational Waves and Gamma-rays from a Binary Neutron
  Star Merger: GW170817 and GRB 170817A}, Astrophys. J. Lett. 848~(2) (2017)
  L13.
\newblock \href {http://arxiv.org/abs/1710.05834} {\path{arXiv:1710.05834}},
  \href {https://doi.org/10.3847/2041-8213/aa920c}
  {\path{doi:10.3847/2041-8213/aa920c}}.

\bibitem{Kouwn:2012np}
S.~Kouwn, P.~Oh, {Dark energy with logarithmic cosmological fluid}, J. Korean
  Phys. Soc. 65~(6) (2014) 814--820.
\newblock \href {http://arxiv.org/abs/1201.4544} {\path{arXiv:1201.4544}},
  \href {https://doi.org/10.3938/jkps.65.814} {\path{doi:10.3938/jkps.65.814}}.

\bibitem{Oikonomou:2019nmm}
V.~K. Oikonomou, {Generalized logarithmic equation of state in classical and
  loop quantum cosmology dark energy\textendash{}dark matter coupled systems},
  Annals Phys. 409 (2019) 167934.
\newblock \href {http://arxiv.org/abs/1907.02600} {\path{arXiv:1907.02600}},
  \href {https://doi.org/10.1016/j.aop.2019.167934}
  {\path{doi:10.1016/j.aop.2019.167934}}.

\bibitem{Chevallier:2000qy}
M.~Chevallier, D.~Polarski, {Accelerating universes with scaling dark matter},
  Int. J. Mod. Phys. D 10 (2001) 213--224.
\newblock \href {http://arxiv.org/abs/gr-qc/0009008}
  {\path{arXiv:gr-qc/0009008}}, \href
  {https://doi.org/10.1142/S0218271801000822}
  {\path{doi:10.1142/S0218271801000822}}.

\bibitem{Linder:2002et}
E.~V. Linder, {Exploring the expansion history of the universe}, Phys. Rev.
  Lett. 90 (2003) 091301.
\newblock \href {http://arxiv.org/abs/astro-ph/0208512}
  {\path{arXiv:astro-ph/0208512}}, \href
  {https://doi.org/10.1103/PhysRevLett.90.091301}
  {\path{doi:10.1103/PhysRevLett.90.091301}}.

\bibitem{Yang:2021flj}
W.~Yang, E.~Di~Valentino, S.~Pan, Y.~Wu, J.~Lu, {Dynamical dark energy after
  Planck CMB final release and $H_0$ tension}, Mon. Not. Roy. Astron. Soc.
  501~(4) (2021) 5845--5858.
\newblock \href {http://arxiv.org/abs/2101.02168} {\path{arXiv:2101.02168}},
  \href {https://doi.org/10.1093/mnras/staa3914}
  {\path{doi:10.1093/mnras/staa3914}}.

\bibitem{Escamilla-Rivera:2021boq}
C.~Escamilla-Rivera, A.~N\'ajera, {Dynamical dark energy models in the light of
  gravitational-wave transient catalogues}, JCAP 03~(03) (2022) 060.
\newblock \href {http://arxiv.org/abs/2103.02097} {\path{arXiv:2103.02097}},
  \href {https://doi.org/10.1088/1475-7516/2022/03/060}
  {\path{doi:10.1088/1475-7516/2022/03/060}}.

\bibitem{Efstathiou:1999tm}
G.~Efstathiou, {Constraining the equation of state of the universe from distant
  type Ia supernovae and cosmic microwave background anisotropies}, Mon. Not.
  Roy. Astron. Soc. 310 (1999) 842--850.
\newblock \href {http://arxiv.org/abs/astro-ph/9904356}
  {\path{arXiv:astro-ph/9904356}}, \href
  {https://doi.org/10.1046/j.1365-8711.1999.02997.x}
  {\path{doi:10.1046/j.1365-8711.1999.02997.x}}.

\bibitem{Christopherson:2008ry}
A.~J. Christopherson, K.~A. Malik, {The non-adiabatic pressure in general
  scalar field systems}, Phys. Lett. B 675 (2009) 159--163.
\newblock \href {http://arxiv.org/abs/0809.3518} {\path{arXiv:0809.3518}},
  \href {https://doi.org/10.1016/j.physletb.2009.04.003}
  {\path{doi:10.1016/j.physletb.2009.04.003}}.

\bibitem{Malik:2008yp}
K.~A. Malik, D.~R. Matravers, {A Concise Introduction to Perturbation Theory in
  Cosmology}, Class. Quant. Grav. 25 (2008) 193001.
\newblock \href {http://arxiv.org/abs/0804.3276} {\path{arXiv:0804.3276}},
  \href {https://doi.org/10.1088/0264-9381/25/19/193001}
  {\path{doi:10.1088/0264-9381/25/19/193001}}.

\end{thebibliography}

\end{document}